\documentclass[preprint,superscriptaddress,showkeys,showpacs,tightenlines,nofootinbib,byrevtex]{revtex4}  
\usepackage{graphicx}
\begin{document}      
\preprint{PNU-NTG-12/2007}
\preprint{YITP-07-60}
\preprint{RCNP-Th07014}
\title{A new nucleon resonance in $\eta$ photoproduction}
%--------------------------------------------------------
\author{Ki-Seok Choi}
\email{kschoi@pusan.ac.kr}
\affiliation{Department of
Physics and Nuclear physics \& Radiation Technology Institute (NuRI),
Pusan National University, Busan 609-735, Korea} 
%--------------------------------------------------------
\author{Seung-il Nam}
\email{sinam@yukawa.kyoto-u.ac.jp}
\affiliation{Yukawa Institute for Theoretical Physics (YITP), Kyoto
University, Kyoto 606-8502, Japan} 
%--------------------------------------------------------
\author{Atsushi Hosaka}
\email{hosaka@rcnp.osaka-u.ac.jp}
\affiliation{Research Center for Nuclear Physics (RCNP), Ibaraki, Osaka
567-0047, Japan}
%--------------------------------------------------------
\author{Hyun-Chul Kim}
\email{hchkim@pusan.ac.kr}
\affiliation{Department of
Physics and Nuclear physics \& Radiation Technology Institute (NuRI),
Pusan National University, Busan 609-735, Korea} 
%--------------------------------------------------------
\date{\today}
%-----------------------------------------------------
\begin{abstract}
We present in this talk recent investigations on the 
nucleon-like resonance $N^*(1675)$ newly found in $\eta$
photoproduction by the GRAAL, Tohoku LNS-$\gamma$ and CB-ELSA
collaborations.  We focus on the production mechanism 
of the $N^*(1675)$, examining its spin and parity theoretically within
the framework of the effective Lagrangian method.  We expliucitly
consider $D_{13}(1520)$, $S_{11}(1535)$, $S_{11}(1650)$,
$D_{15}(1675)$, $P_{11}(1710)$, $P_{13}(1720)$ as well as possible
background contributions.  We calculate the differential cross
sections and beam asymmetries for the neutron and proton targets. It
turns out that there is manifest isospin asymmetry in $\eta$
photoproduction, which can be explained by the asymmetry in the
transition magnetic moments: $\mu_{\gamma{p}p^*} \ll
\mu_{\gamma{n}n^*}$.  Moreover, we find that the spin-1/2 state is  
preferred and this observation implies that the new resonance may be
identified as a non-strangeness member of the baryon antidecuplet.     
\end{abstract} 
%-------------------------------------------------------
\maketitle
%-------------------------------------------------------
\section{Introduction}
%-------------------------------------------------------
Recently, the GRAAL collaboration has reported a new nucleon resonance
$N^*(1675)$ from $\eta$
photoproduction~\cite{Kuznetsov:2004gy,Kuznetsov:2006kt},  
which has a narrow decay width: $\Gamma_{N^*\to\eta N}\simeq 40$ MeV.
The Fermi-motion corrections being taken into account, 
the width may become even narrower: $\sim10$ MeV~\cite{Fix:2007st}. 
This narrow width is a typical feature for the pentaquark exotic
baryons~\cite{Diakonov:2003jj,Arndt:2003ga,Polyakov:2003dx}.
Moreover, the production process of the $N^*(1675)$ largely depends on
its isospin state of the target nucleons: A larger $N^*(1675)$ peak is
observed for the neutron target, while it is suppressed for the proton 
one, i.e. $\mu_{\gamma nn^*}\gg\mu_{\gamma pp^*}$.  Interestingly,
$N^*(1675)$ being assumed as a member of the baryon antidecuplet
($\overline{10}$), this large isospin asymmetry can be well explained
in the chiral quark-soliton model
($\chi$QSM)~\cite{Kim:2005gz,Azimov:2005jj}.  In fact, the strong
isospin asymmetry is an exact consequence of $U$-spin conservation in
the flavor SU(3) limit~\cite{Hosaka}.  Concerning the spin and parity,
their assignments are not yet determined unambiguously. Although the
$\eta$-MAID has assumed $J^P=1/2^+$ as suggested by the
$\chi$QSM~\cite{Fix:2007st}, in our previous work~\cite{Choi:2005ki},
we have shown that $J^P=1/2^-$ is equally possible in comparison with
the experimental data. In the present work, we want to investigate the
$\eta$ photoproduction, including the following six nucleon
resonances, $D_{13}(1520)$, $S_{11}(1535)$, $S_{11}(1650)$,  
$D_{15}(1675)$, $P_{11}(1710)$ and $P_{13}(1720)$, in a fully
relativistic manner.  We ignore the contributions from $N^*(1680)$ and 
$N^*(1700)$ considered in Ref.~\cite{Fix:2007st}, since 
their branching ratios to $\eta N$ channel are negligible.  The
nucleon-pole terms and vector-meson exchanges are also taken into  
account as backgrounds. In order to test the spin and parity of the
new resonance, we investigate four different cases:
$J^P=1/2^{\pm},3/2^{\pm}$.  As a result, we observe
that $\mu_{\gamma nn^*(1675)}=0.1\sim0.2$ and $\mu_{\gamma
  pp^*(1675)}\simeq0$ for $J^P=1/2^{\pm}$ whereas $\mu_{\gamma 
nn^*(1675)}=0.01\sim0.02$ and $\mu_{\gamma pp^*(1675)}\simeq0$ for
$J^P=3/2^{\pm}$ to reproduce the GRAAL data qualitatively. However, it
is rather difficult to see obvious peak structures at
$E_{\rm{cm}}\simeq1675$ MeV for the spin-3/2 cases due to the strong
interference with $D_{15}(1675)$.  This observation tells us that the
new resonance may be identified as a member of the baryon antideucplet 
apart from its parity, though there is still a possibility to explain 
it as one of coupled-channel effects near the $K\Sigma$ and $K\Lambda$ 
thresholds.  
%--------------------------------
\section{General formalism}
\label{sec:1}
%--------------------------------
In this Section, we set up a theoretical formalism for the investigation
of the $N^*$ resonance in $\eta$ photoproduction. The effective
Lagrangians for each vertex for the backgrounds can be written as
follows:   
\begin{eqnarray}
 \mathcal{L}_{\gamma NN}
&=&-e_N\bar{N}\gamma_{\mu}NA^{\mu} + 
\frac{ie_Q\kappa_N}{4M_N}\bar{N}\sigma_{\mu\nu}NF^{\mu\nu}+{\rm h.c.},
\nonumber\\
\mathcal{L}_{\eta NN}&=&-ig_{\eta NN}\bar{N}\gamma_5 \eta N+{\rm h.c.},
\nonumber \\
\mathcal{L}_{VNN}&=& -g^v_{VNN} \bar{N}\gamma_{\mu}NV^{\mu}+
 \frac{ig^t_{VNN}}{4M_N}\bar{N}\sigma_{\mu\nu}V^{\mu\nu}N+{\rm h.c.}.,
\nonumber\\
\mathcal{L}_{ \gamma\eta V}&=&\frac{e_Qg_{ \gamma\eta V}}{4M_{\eta}}
\epsilon_{\mu\nu\sigma\rho}F^{\mu\nu}V^{\sigma\rho}\eta+{\rm h.c.}, 
\label{backgrounds}
\end{eqnarray}
where $\gamma$, $N$, $\eta$ and $V$ stand for the fields of the
photon, nucleon, $\eta$ meson and vector mesons ($\rho$ and $\omega$), 
respectively. $e_N$ and $\kappa_N$ denote the electric charge and
anomalous magnetic moment of the nucleon, respectively, while $e_Q$
the unit charge.  Generically, the $M_h$ denotes the mass of the
hadron $h$. The strength of the meson-baryon coupling constants are
employed from the Nijmegen potential~\cite{Stoks:1999bz} and given in
our previous work~\cite{Choi:2007gy}.  In the following, we present
the effective Lagrangians for the resonant contributions of spin $1/2$,
$3/2$ and $5/2$: 
\begin{eqnarray}
{\cal L}_{\gamma N N^*}^{1/2}&=& \frac{\mu_{\gamma N N^*}
 }{2(M_N+M_{N^*})}\bar{N}^*\Gamma^a_5\sigma_{\mu\nu}F^{\mu\nu}N,
\nonumber\\ 
 {\cal L}_{\eta N N^*}^{1/2}&=&-ig_{\eta N N^*} \bar{N }
\Gamma^a_5 \gamma_5 \eta N^*,
\nonumber\\
{\cal L}_{\eta N N^*}^{3/2}&=&\frac{g_{\eta N
    N*}}{M_{\eta}}\bar{N}^{*\mu}\Theta_{\mu\nu}(A,B)\Gamma^a_5 N,
    \partial^{\nu}\eta \nonumber \\
{\cal L}_{\gamma N
    N^*}^{3/2}&=&\frac{i\mu_{\gamma{N}N^*}}{M_{N^*}}\bar{N}^{*\mu}
\Theta_{\mu\nu}(C,D)\Gamma^b_5\gamma_{\lambda}NF^{\lambda \nu},
\nonumber\\
{\cal L}_{\eta N N^*}^{5/2}&=&\frac{g_{\eta N
    N^*}}{m^2_{\eta}}\bar{N}^{*\mu\nu}\Theta_{\mu\delta}(A,B)
\Theta_{\nu\lambda}(C,D)\Gamma^b_5N
    \partial^{\delta}\partial^{\lambda}\eta, 
\nonumber\\
{\cal L}_{\gamma N
    N^*}^{5/2}&=&\frac{\mu_{\gamma{N}N^*}}{M^2_{N^*}}\bar{N}^{*\mu\alpha}
\Theta_{\mu\nu}(E,F)\gamma_{\lambda}\Gamma^a_5(\partial_{\alpha}
F^{\lambda\nu})N,
\label{resonance}
\end{eqnarray}
where the spinors for spin-3/2 and spin-5/2 fermions are defined by
the Rarita-Schwinger formalism~\cite{Titov:2002iv,Rarita:1941mf}. For
convenience, we turn off the off-shell factor in
$\Theta_{\mu\nu}(A,B)$, since its effects are rather small.  Being
similar to the background contributions, the strong couplings can be
determined by two-body decay process with the Yukawa interactions
given in Eq.~(\ref{resonance}).  Physical inputs (full decay width and 
branching ratio for each resonance) are also given in
Ref.~\cite{Choi:2007gy}. We note that the values of the physical
inputs are compatible with those given by the PDG~\cite{Yao:2006px}.
The transition photon couplings ($\mu_{\gamma NN^*}$) can be computed
via their helicity amplitudes~\cite{Titov:2002iv}. The invariant
amplitudes for each contribution can be evaluated straightforwardly  
by using Eqs.~(\ref{backgrounds}) and (\ref{resonance}) as shown in
Ref.~\cite{Choi:2007gy}.    
%--------------------------------
\section{Numerical results}
\begin{figure}[t]
\begin{tabular}{cccc}
\includegraphics[width=4cm]{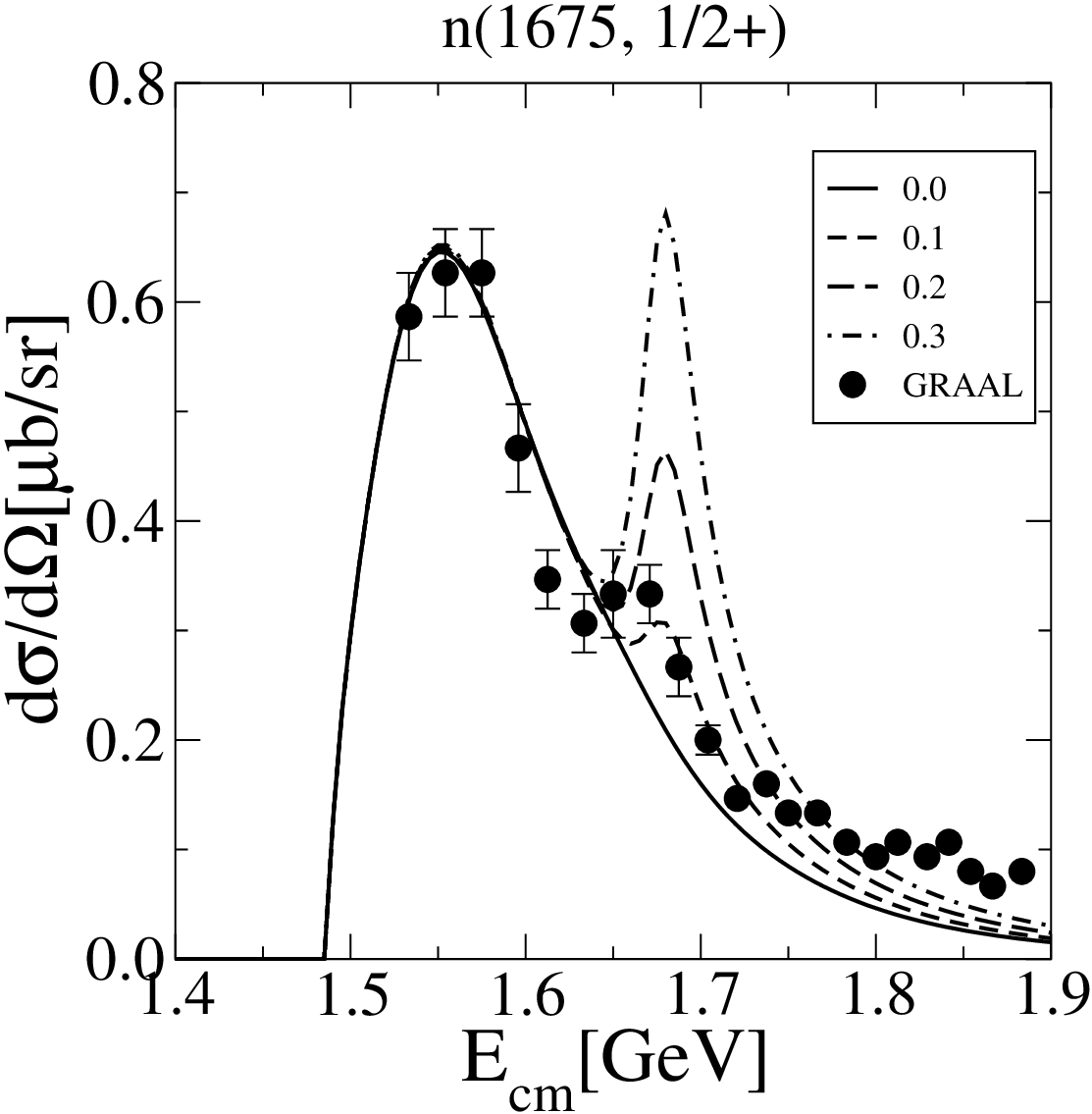}
\includegraphics[width=4cm]{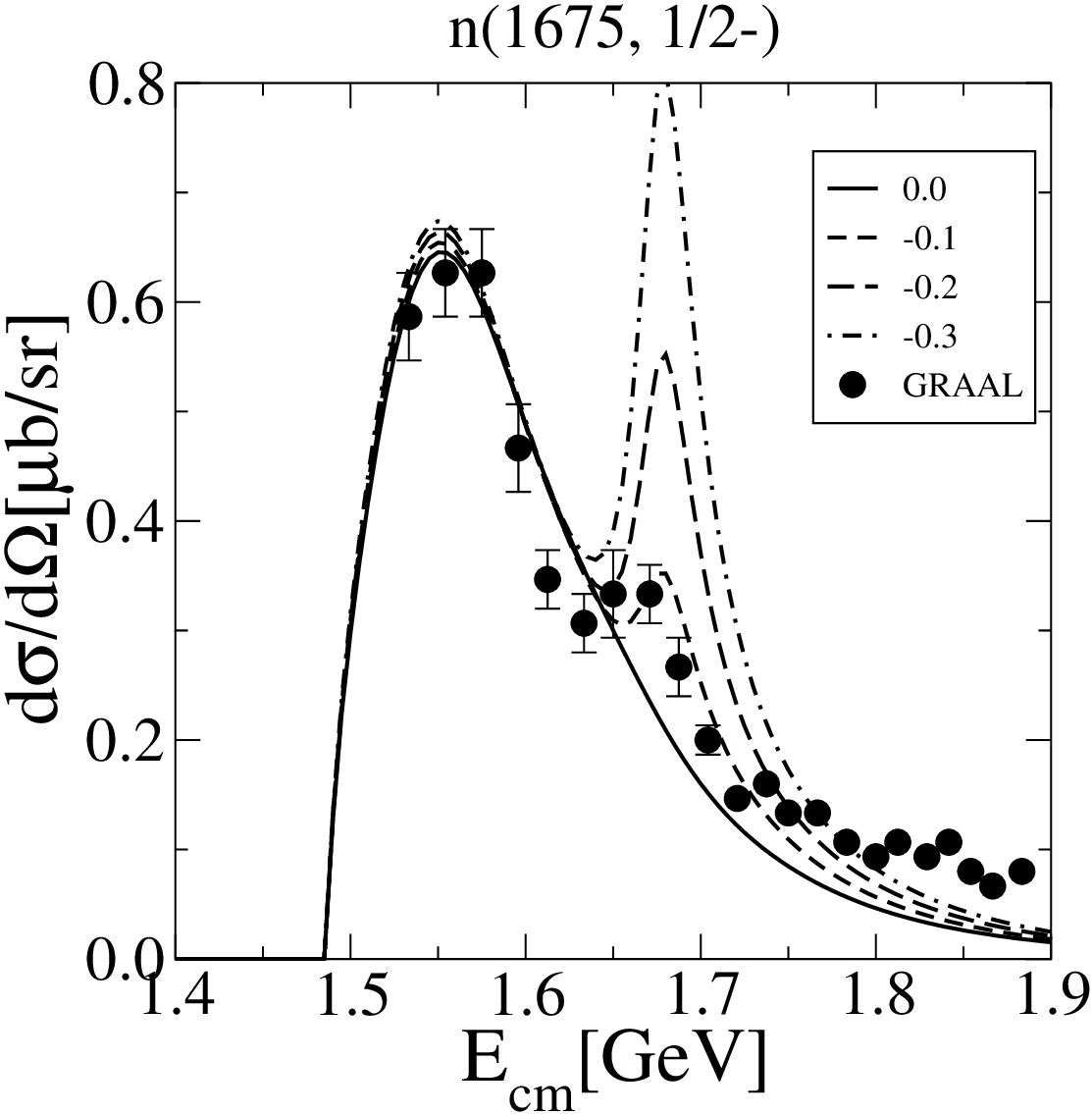}
\includegraphics[width=4cm]{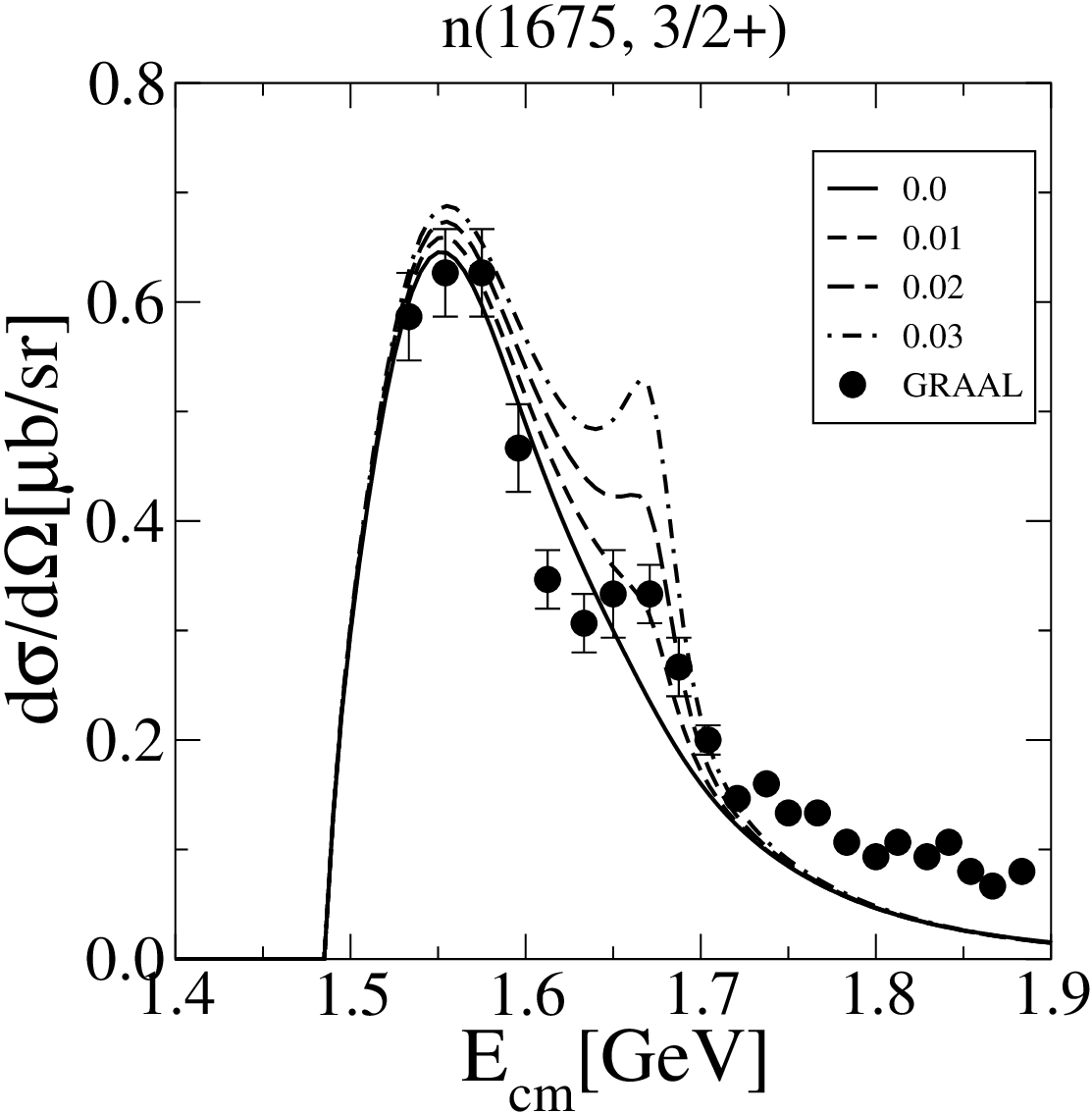}
\includegraphics[width=4cm]{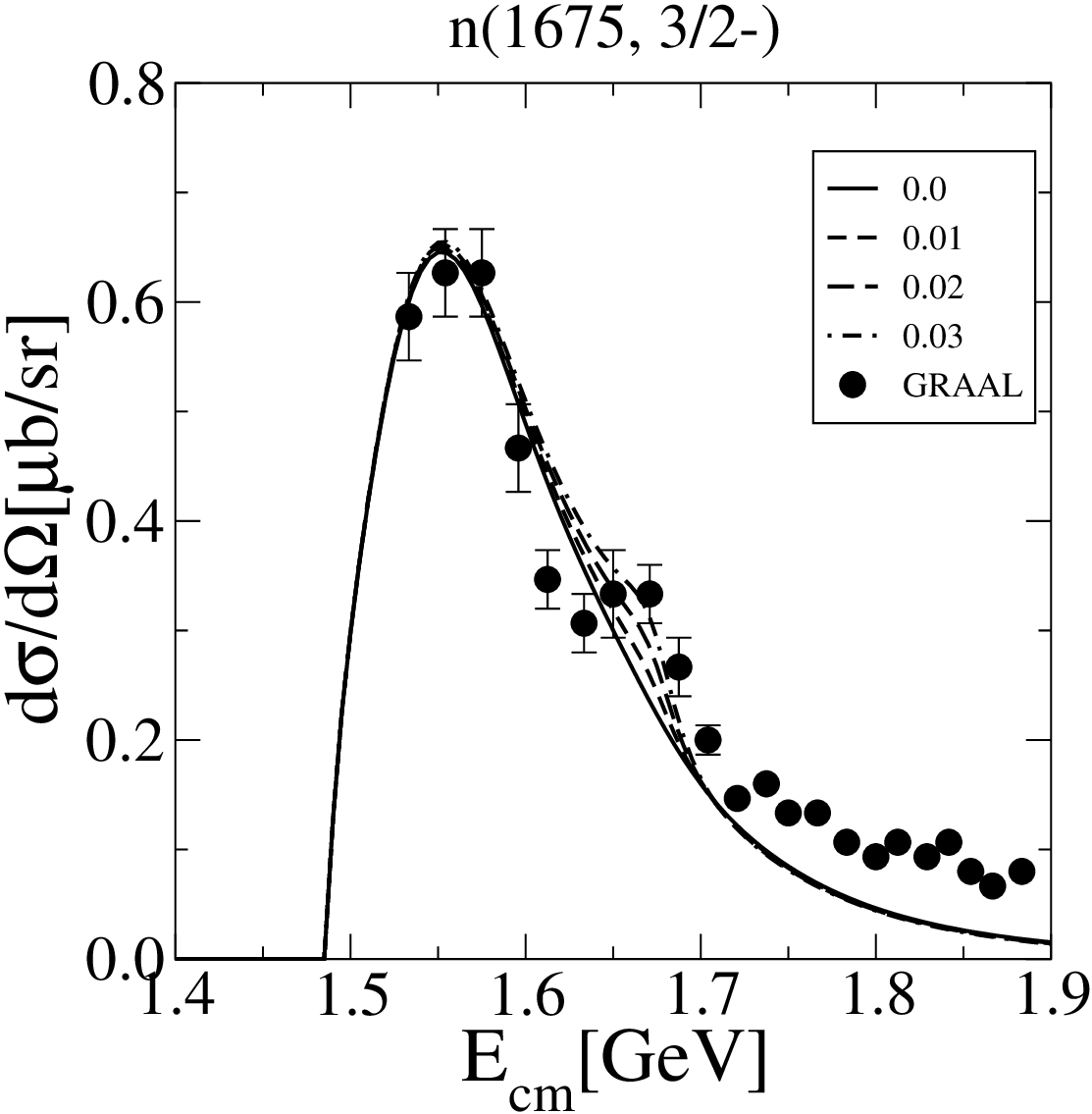}
\end{tabular}
\begin{tabular}{cccc}
\includegraphics[width=4cm]{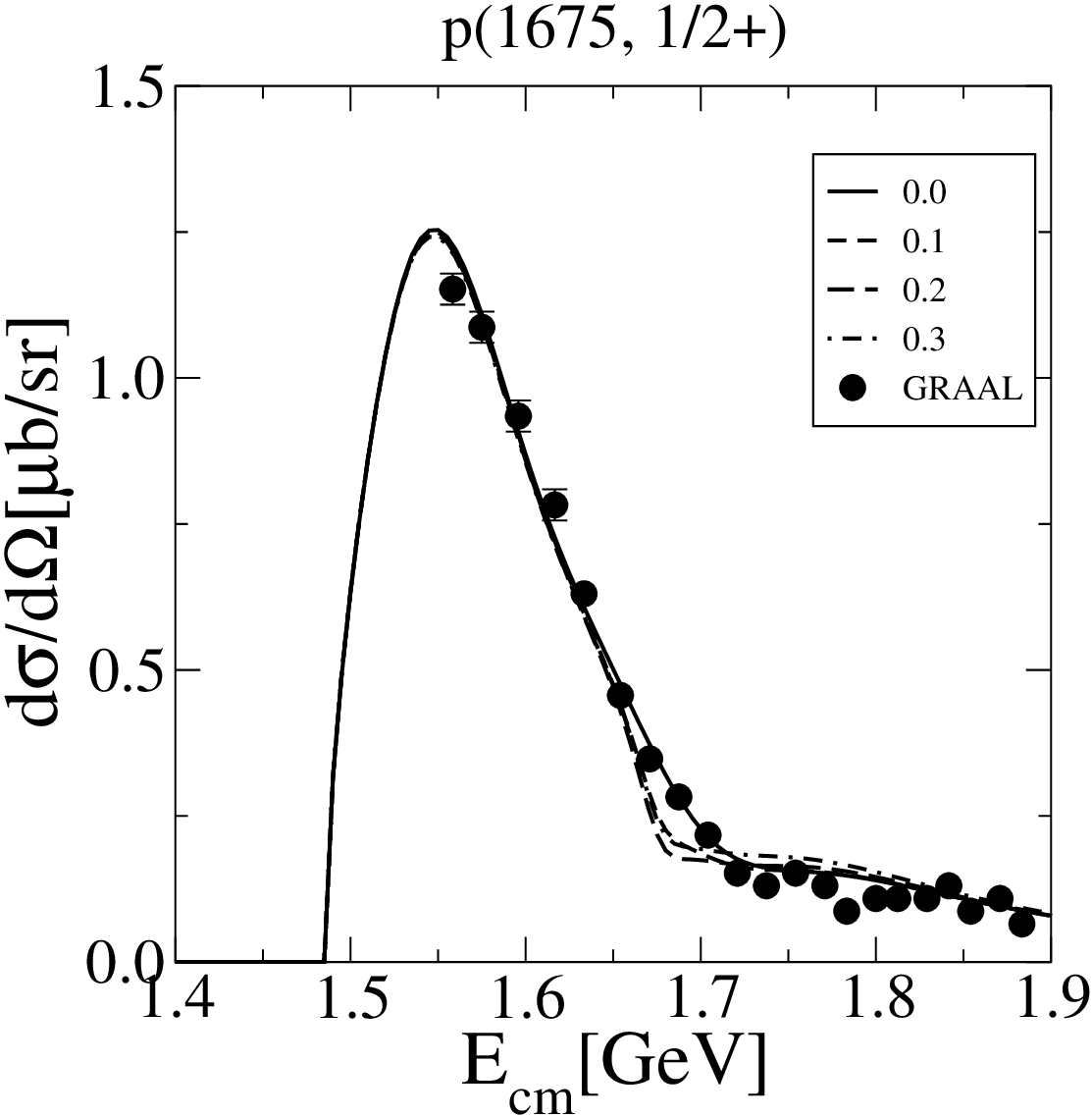}
\includegraphics[width=4cm]{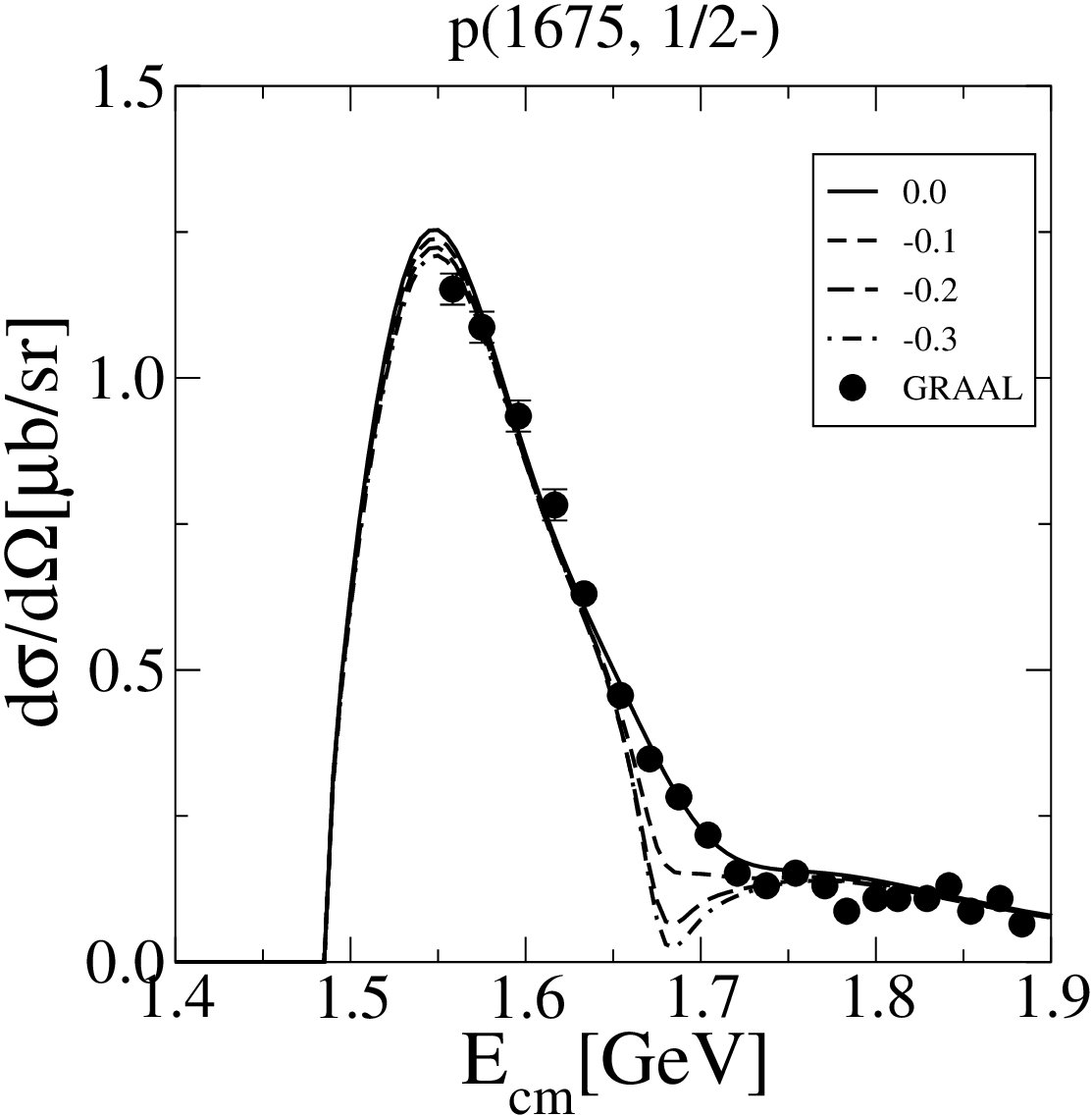}
\includegraphics[width=4cm]{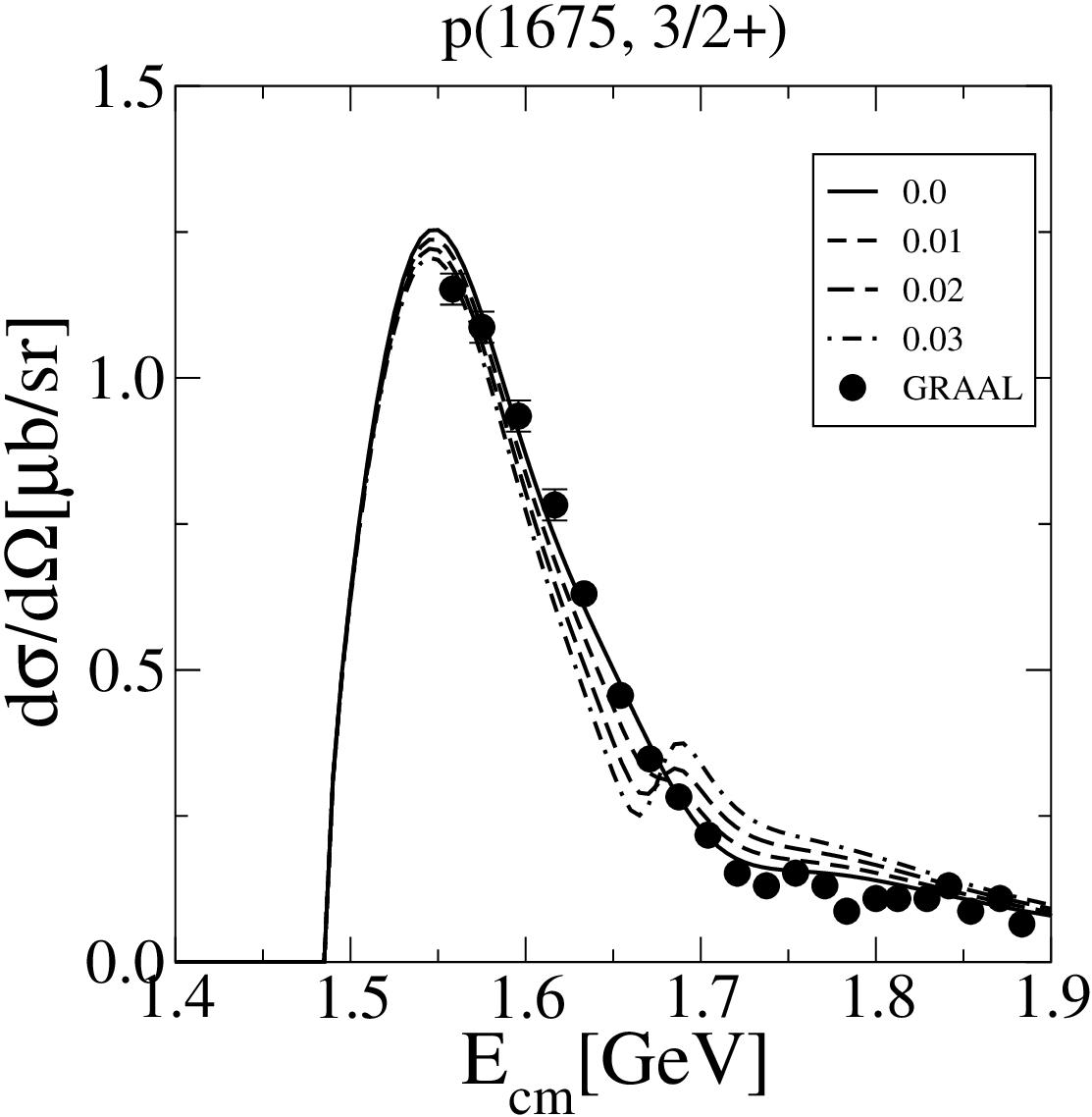}
\includegraphics[width=4cm]{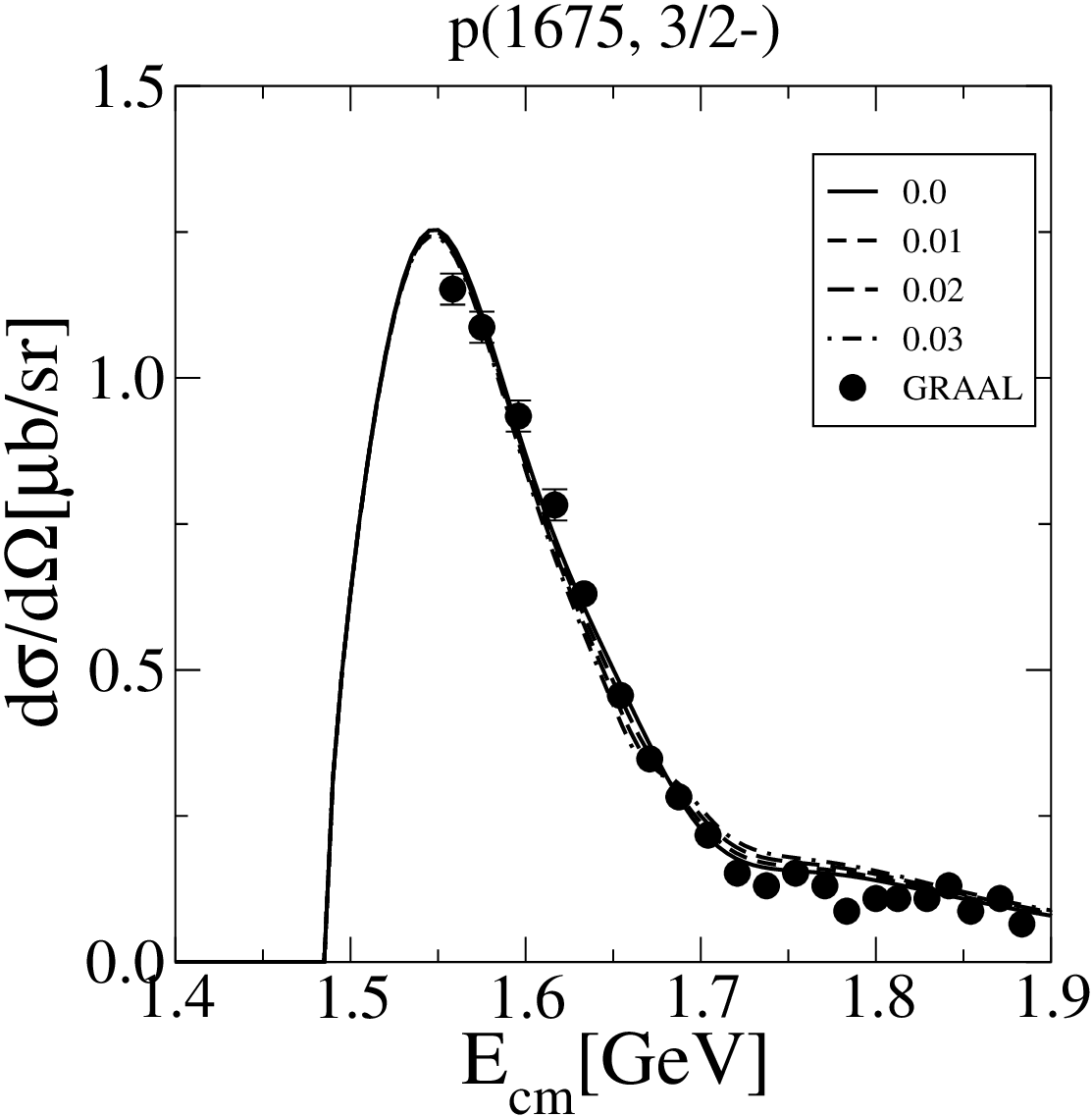}
\end{tabular}
\caption{Differential cross sections for $\eta$ photoproduction at
  $\theta_{\gamma\eta}=140^{\circ}$ for the neutron (upper panels) and
  proton (lower panels) targets. Spin and parity of $N^*(1675)$ are
  assigned by $1/2^+$, $1/2^-$, $3/2^+$ and $3/2^-$ from the left to
  the right. Experimental data are taken from
  Ref.~\cite{Kuznetsov:2004gy,Kuznetsov:2006kt}.}  
\label{fig1}
\end{figure}

\begin{figure}[t]
\begin{tabular}{cccc}
\includegraphics[width=4cm]{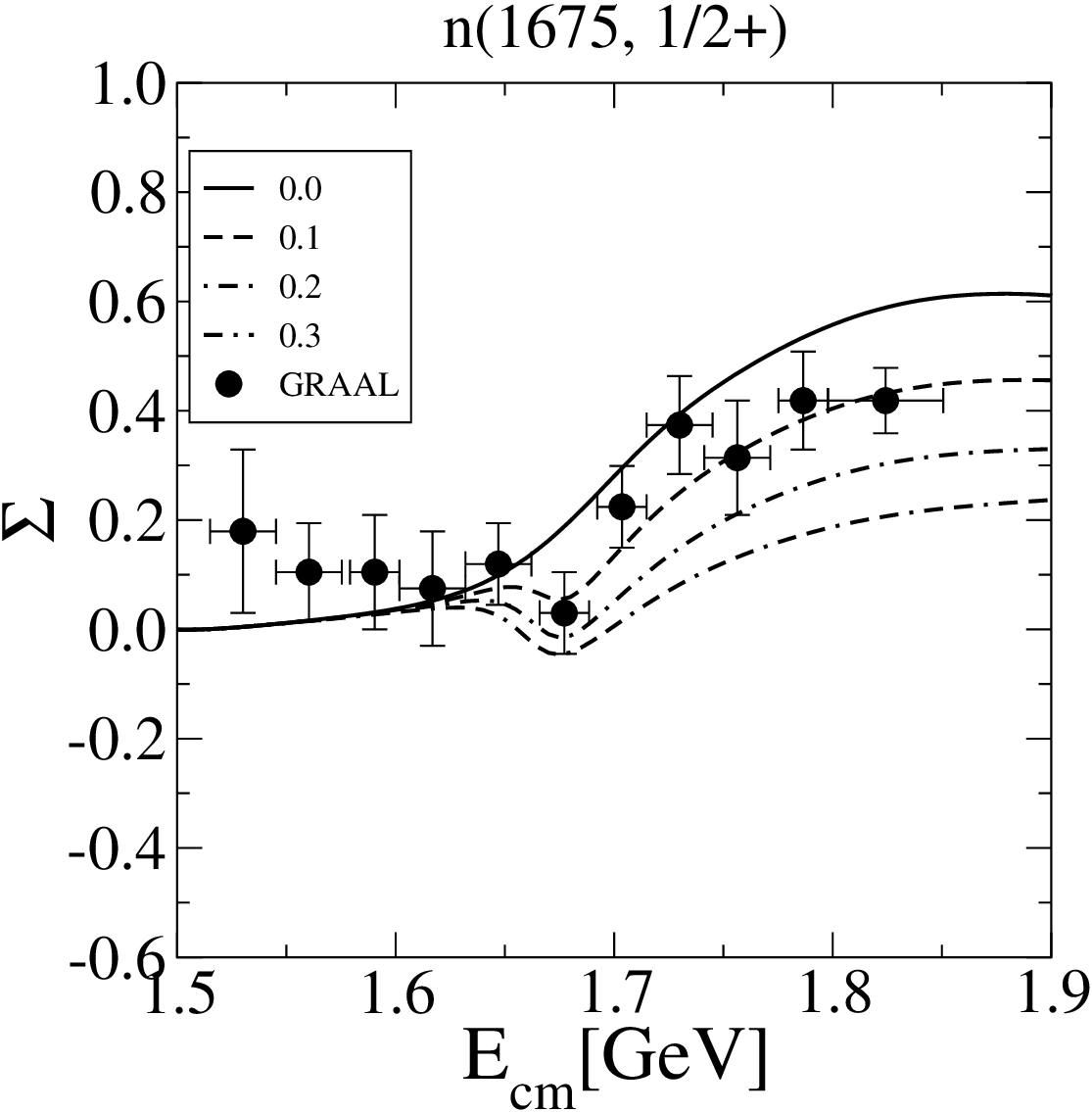}
\includegraphics[width=4cm]{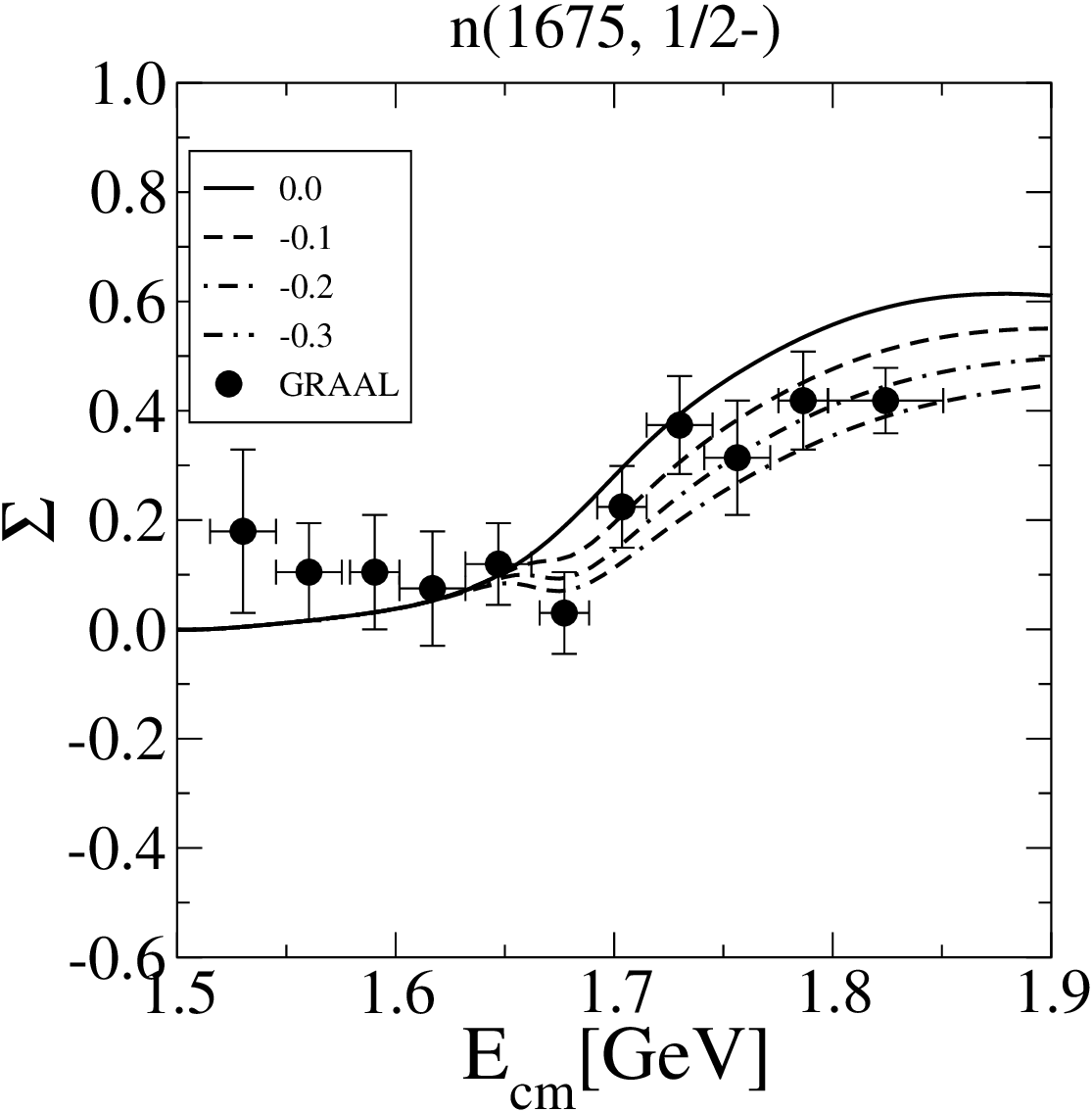}
\includegraphics[width=4cm]{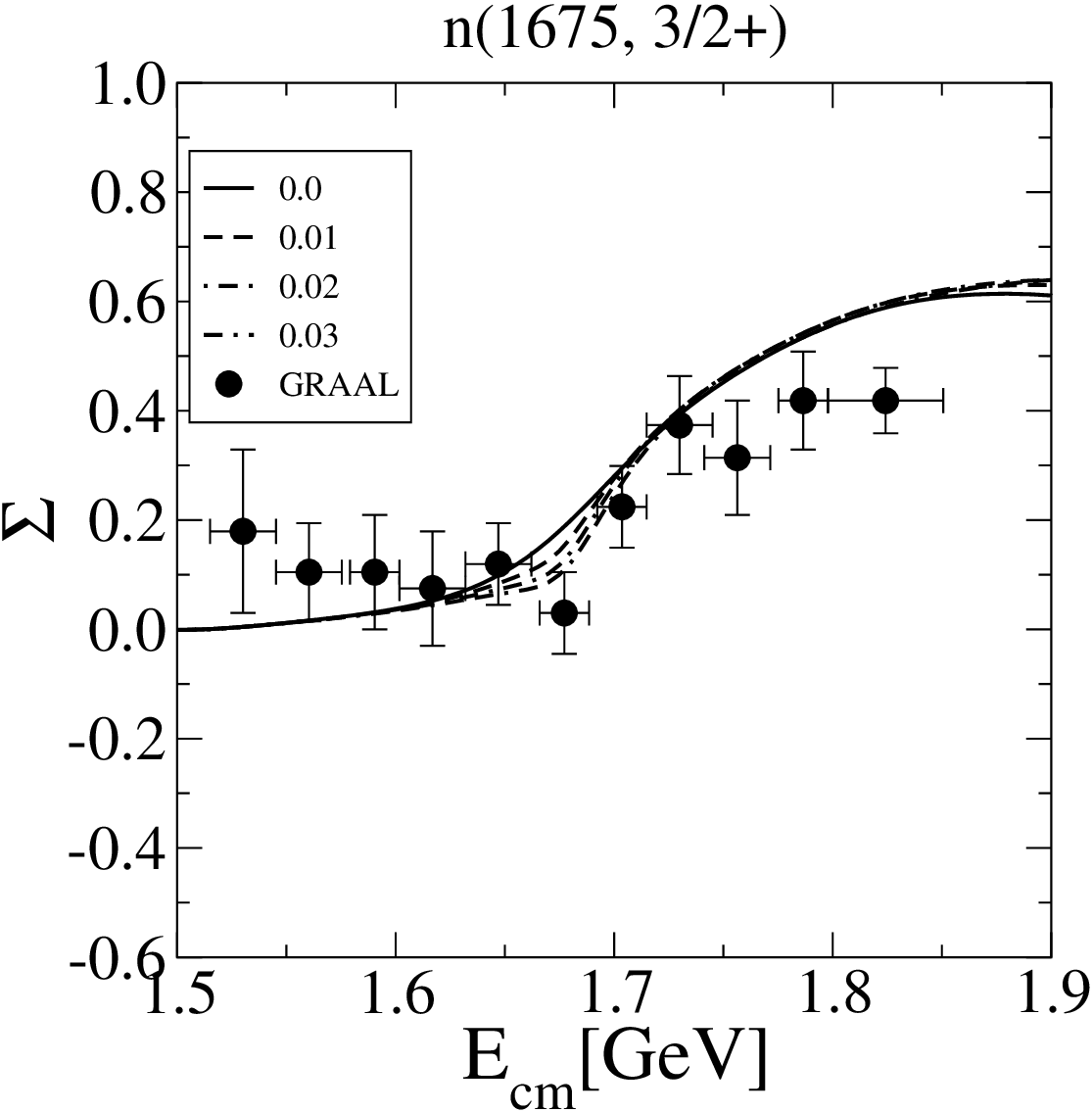}
\includegraphics[width=4cm]{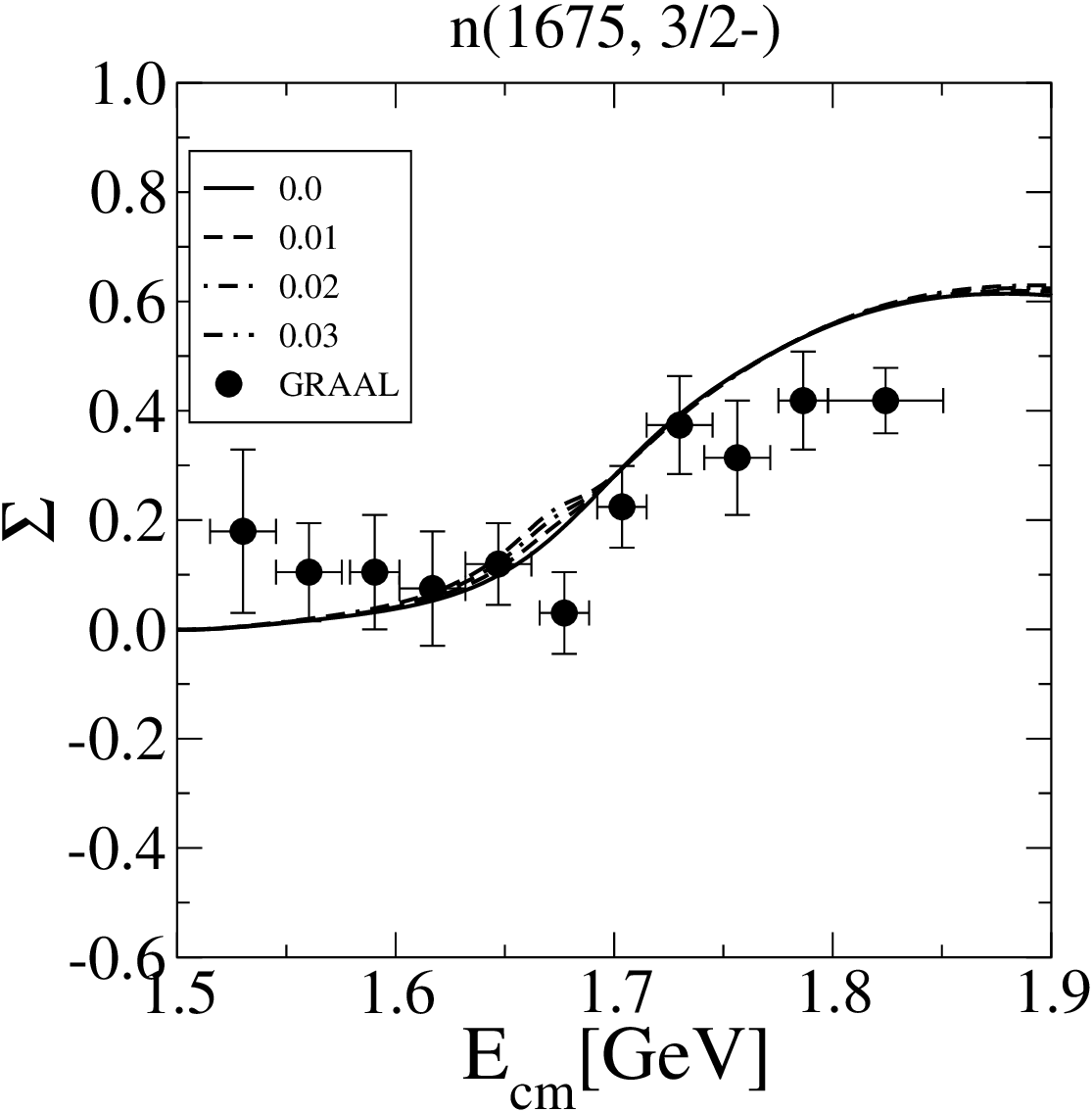}
\end{tabular}
\begin{tabular}{cccc}
\includegraphics[width=4cm]{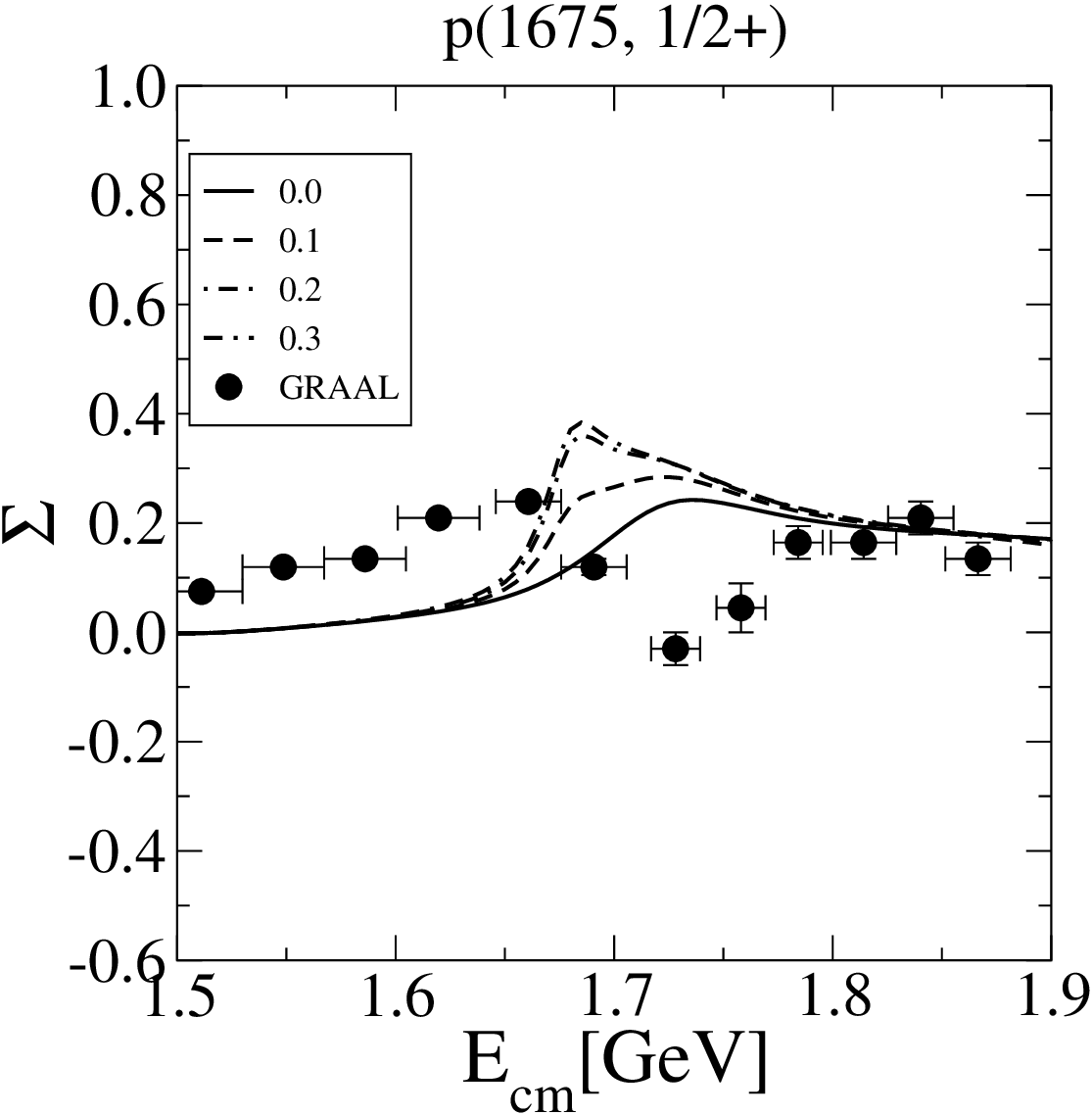}
\includegraphics[width=4cm]{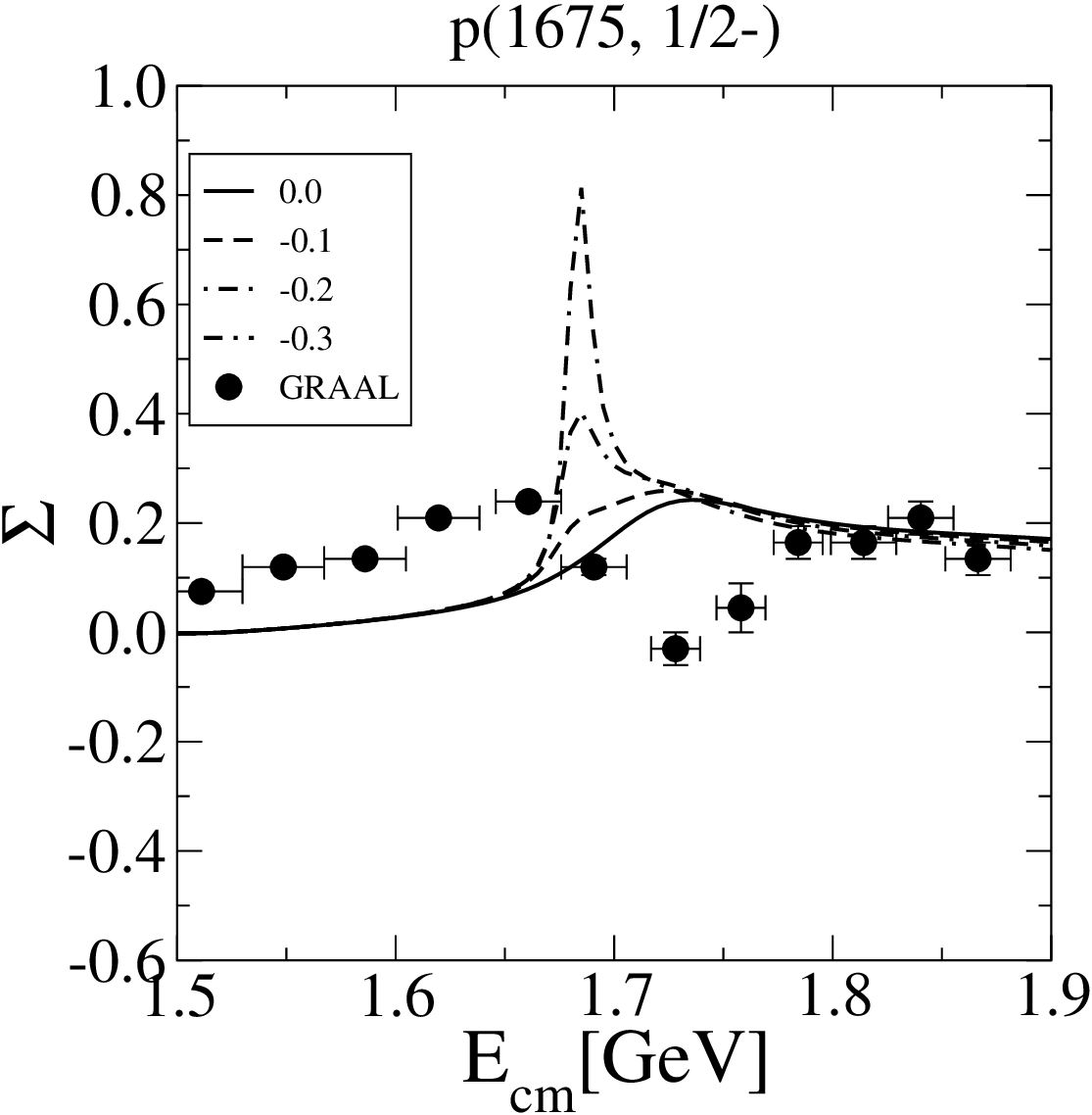}
\includegraphics[width=4cm]{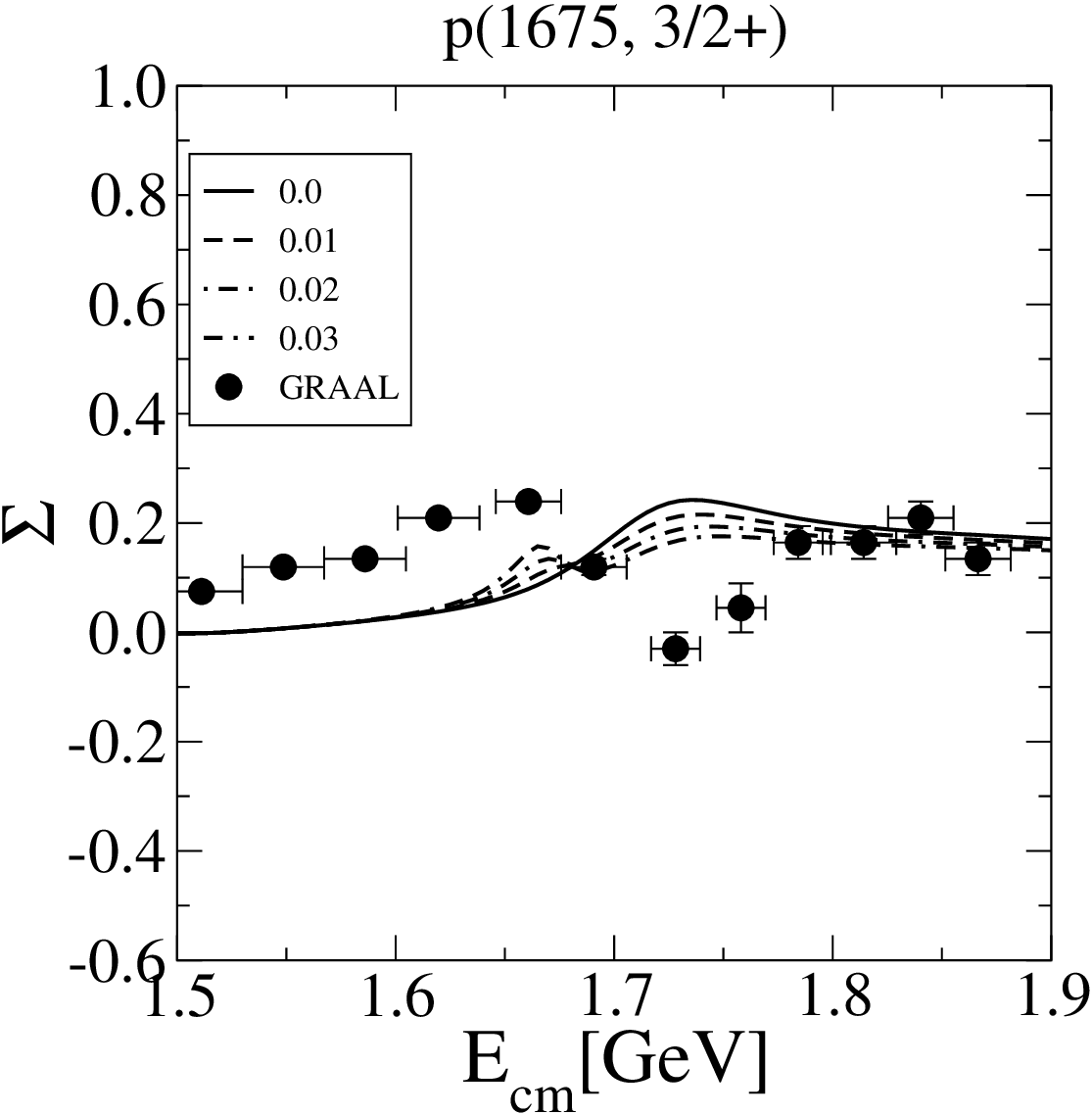}
\includegraphics[width=4cm]{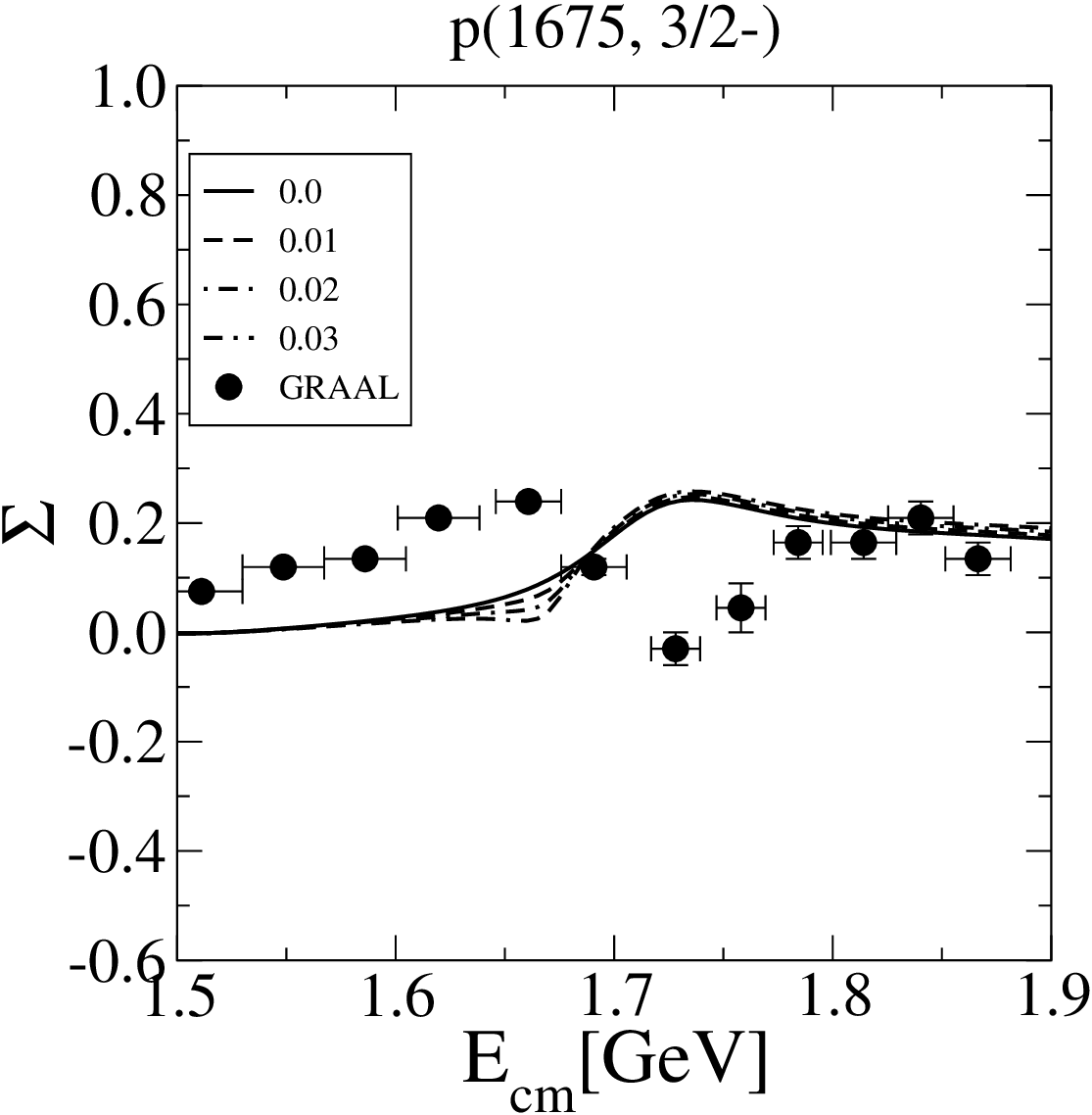}
\end{tabular}
\caption{Beam asymmetries for $\eta$ photoproduction for the neutron
  (upper row) and proton (lower row) targets at
  $\theta_{\gamma\eta}=140^{\circ}$. Spin and parity of $N^*(1675)$
  are assigned by $1/2^+$, $1/2^-$, $3/2^+$ and $3/2^-$ from the left
  to the right. Experimental data are taken from
  Ref.~\cite{Kuznetsov:2004gy,Kuznetsov:2006kt}.}  
\label{fig2}
\end{figure}

%--------------------------------
We now show the numerical results for differential cross
sections with respect to the center of mass energy ($E_{\rm cm}$) in
Figure~\ref{fig1} for the neutron (upper panels) and proton (lower
panels) targets. All curves are drawn at $\theta_{\gamma\eta} =
140^{\circ}$, which is the angle between the incident photon and
outgoing $\eta$ meson.  The spin and parity of $N^*(1675)$ are
assigned as $1/2^+$, $1/2^-$, $3/2^+$ and $3/2^-$ from the left to the
right.  We plot the results for different strengths of the transition
photon couplings, $\mu_{\gamma NN^*(1675)}$.  As for the neutron
target, we can reproduce the data qualitatively 
well as shown in the upper panel of Figure~\ref{fig1}. In the vicinity
of the threshold, the $S_{11}(1535)$ contribution dominates.  The
numerical results are underestimated for the region beyond
$E_{\rm{cm}}\sim1.8$ GeV due to the destructive interference between
$S_{11}(1535)$ and the vector mesons, and to the weak $D_{15}(1675)$
contribution ($\Gamma_{D_{15}(1675)}/\Gamma_{D_{15}(1675)\to\eta
  N}\simeq1\%$) which is similar to those in the model (ii) of
Ref.~\cite{Fix:2007st}.   We observe sharp peak structures at
$E_{\rm{cm}}\sim1.675$ GeV for $J^P=1/2^{\pm}$. Comparing them to the
experimental data taken from the GRAAL
experiment~\cite{Kuznetsov:2004gy,Kuznetsov:2006kt}, we estimate 
the transition magnetic moment for the $n^*\to n\gamma$ decay:
$|\mu_{\gamma  nn^*(1675,1/2^{\pm})}|=0.1\sim0.2$ for the spin 1/2
case. On the contrary to the spin-1/2 cases, it is rather difficult to
find obvious peak structures for the spin-3/2 case for both parities, 
because of the strong destructive interference with $D_{15}(1675)$.
Note that the strength of the $\mu_{\gamma nn^*(1675,3/2^{\pm})}$
turns out to be around ten times as small as those of the spin-1/2
case.  The reason for this smallness lies in the fact that the higher
partial-wave contributions come into play in the case of the spin-3/2
$N^*$. Now we are in a position to discuss the results of the proton target
shown in the lower panel of Figure~\ref{fig1}.  The value of
$|\mu_{\gamma pp^*(1675)}|$ should be nearly zero to produce the experimental
curves for all spin and parity states.  In contrast to the neutron
target, we can produce the data relatively well for the regions above
$\sim1.8$ GeV due to the constructive interference between
$S_{11}(1535)$ and the vector mesons. From the numerical results for
the differential cross sections, we observe that the new resonance has
strong isospin asymmetry by employing the transition photon couplings:
$|\mu_{\gamma nn^*(1675)}|=0.1\sim0.2$ and $|\mu_{\gamma
  pp^*(1675)}|\simeq0$ for the spin-1/2 case as discussed 
already in the previous work~\cite{Choi:2005ki}.  However, it turns out 
that no clear peak structure is shown for the spin-3/2 $N^*$.  We
emphasize that all of these results support the prediction for
$|\mu_{\gamma NN^*(1675)}|$ from the $\chi$QSM~\cite{Kim:2005gz} in
which this new resonance $N^*$ is treated as a non-strangeness member
of the baryon antidecuplet.  From the  results discussed above,
however, we are not able to determine the parity of the resonance
within the present theoretical framework, since the results for the
$1/2^-$ $N^*$ are also compatible with the data.  It is worthwhile
to mention that the $J^P=1/2^-$ baryon state can not be 
achieved by the conventional collective quantization of the soliton in
the $\chi$QSM.     
We now present the numerical results for the beam asymmetries
as a function of $E_{\rm cm}$, which plays an essential role in 
determining the new
resonance~\cite{Kuznetsov:2004gy,Kuznetsov:2006kt}.  First, we define 
the beam asymmetry as
follows~\cite{Kuznetsov:2004gy,Kuznetsov:2006kt}:  
\begin{equation} 
\Sigma=\left[\frac{d\sigma}{d\Omega}_{\parallel}
-\frac{d\sigma}{d\Omega}_{\perp}\right]\times\left[\frac{d\sigma}
{d\Omega}_{\parallel}+\frac{d\sigma}{d\Omega}_{\perp}\right]^{-1},
\label{eq:BA}
\end{equation}
where the subscript $\parallel$ denotes the polarization vector of
the incident photon parallel to the reaction plane, and $\perp$ for
the longitudinally polarized photon. Figure~\ref{fig2} depicts the
numerical results for the beam asymmetries as in the same way as
Fig.~\ref{fig1}.  First, we discuss the neutron case as shown in the
upper panel of Fig.~\ref{fig2}.  The results reproduce relatively well
the data except for the region below $\sim1.6$ GeV due to the strong
magnetic $\omega$-meson contribution.  Note that the $\Sigma$ has a
valley structure around $E_{\rm{cm}}\sim1.675$ GeV for the spin-1/2
cases as the value of $|\mu_{\gamma nn^*(1675)}|$ increases.  As
argued in Refs.~\cite{Kuznetsov:2004gy,Kuznetsov:2006kt}, this
tendency can be 
an indication of the new resonance contribution.  Although we still
have some visible structures for the spin-3/2 cases around
$E_{\rm{cm}}\sim1.675$ GeV, they look much weaker than those for the
spin-1/2 cases. In contrast to the neutron case, as seen in the lower
panel of Figure~\ref{fig2}, we fail to reproduce the data for all spin 
and parity states, although order of magnitude is rather
compatible.  Especially, the downward hump and valley structures in
the range of $1.7\,{\rm{GeV}} \le{E_{\rm{cm}}} \le
1.8{\rm{GeV}}$ in the experimental data are not generated at all.  As
the strength of $\mu_{\gamma{p}p^*}$ increases, we observe clear kinks
around $E_{\rm{cm}}=1.7$ GeV for the spin-1/2 cases due to the
interference between the new resonance and $D_{15}(1675)$.  
%--------------------------------
\section{Summary and conclusion}
%--------------------------------
In the present talk, we have shown the recent results for the $\eta$
photoproduction off the nucleon with the effective Lagrangian approach
in the Born approximation.  We focussed on the new nucleon resonance
around $E_{\rm{cm}} \sim 1.675$ GeV, examining its possible spin and
parity theoretically and considering $1/2^{\pm}$ as well as
$3/2^{\pm}$.  First, we observed fine peak structures for the new
resonance in the differential cross sections being compatible with the
GRAAL data, when its spin and parity were assigned to be
$J^P=1/2^{\pm}$. We obtained $|\mu_{\gamma nn^*(1675,1/2^{\pm})}| = 
0.1\sim0.2$ whereas $|\mu_{\gamma  pp^*(1675,1/2^{\pm})}|$ should be
almost zero to meet the data.  In contrast, there were no clear peaks
shown for both $3/2^{\pm}$ due to the strongly destructive
interference with $D_{15}(1675)$ within the present theoretical
framework. From these observations, the new resonance may be
considered as $N^*(1675,1/2^{\pm})$ in the present work. If this is
the case, the resonance may possibly be identified as a
non-strangeness member of the baryon antidecuplet with $J^P=1/2^+$. In
addition, this can be a natural consequence of the $U$-spin symmetry
in the SU(3) baryon representations.  Though the
present work favors the new nucleon resonance $N^* (1675)$ 
as a member of the baryon antidecuplet, we have to mention that it is
necessary to consider a possibility that it might arise from the 
coupled-channel effects near the $K\Sigma$ and $K\Lambda$ thresholds.  

%-------------------------------------------------------
\section*{Acknowledgement}
%-------------------------------------------------------
K.S.C. is gratefule to the organizers for the International Workshop
NSTAR2007, which was held during 5-8 September, 2007 in Bonn,
Germany.  Authors appreciate the fruitful discussions with
V. Kuznetsov.  The present work is supported by the Korea Research
Foundation Grant funded by the Korean Government (MOEHRD)
(KRF-2006-312-C00507).  The work of K.S.C. is partially supported by the
Brain Korea 21 (BK21) project in Center of Excellency for Developing
Physics Researchers of Pusan National University, Korea. The work of
S.i.N. is supported in part by grant for Scientific Research (Priority
Area No.17070002) from the Ministry of Education, Culture, Science and
Technology, Japan. A.H. is supported in part by the Grant for
Scientific Research ((C) No.19540297) from the Ministry of Education,
Culture, Science and Technology, Japan. 
%--------------------------------

\end{document}